\documentclass[runningheads]{llncs}

\usepackage{xspace}
\usepackage{tikz,pifont}
\usepackage{amsmath,amssymb,amsfonts}
\usepackage{listings}
\usepackage{algorithmic}
\usepackage{graphicx}
\usepackage{textcomp}
\usepackage{xcolor, colortbl}
\usepackage{makecell}
\usepackage{hyperref}
\usepackage{xcolor}
\usepackage{amssymb}
\usepackage[many]{tcolorbox}

\usepackage{enumitem}
\newlist{questions}{enumerate}{2}
\setlist[questions,1]{label=\textbf{RQ\arabic*.},ref=\textbf{RQ\arabic*}}
\setlist[questions,2]{label=(\alph*),ref=\thequestionsi(\alph*)}

\newcommand*{\ie}{i.e.,\@\xspace}

\makeatother

\definecolor{verylightgray}{gray}{0.92}
\definecolor{ao(english)}{rgb}{0.0, 0.5, 0.0}

\definecolor{deepblue}{rgb}{0,0,0.5}
\definecolor{deepred}{rgb}{0.6,0,0}
\definecolor{deepgreen}{rgb}{0,0.5,0}
\definecolor{shadecolor}{gray}{0.9}

\colorlet{shadecolor}{verylightgray}
\colorlet{framecolor}{black}

\lstdefinestyle{searchstringstyle}{
	basicstyle=\ttfamily\scriptsize,
	breaklines=true,
	captionpos=b,
	numbers=none,
	numbersep=4pt,
	showspaces=false,
	showstringspaces=false,
	showtabs=false,
	tabsize=2,
	frame=single
}

\definecolor{mygreen}{rgb}{0,0.6,0}
\definecolor{mygray}{rgb}{0.95,0.95,0.95}
\definecolor{myred}{rgb}{0.5,0,0}

\definecolor{verylightgray}{gray}{0.92}
\definecolor{ao(english)}{rgb}{0.0, 0.5, 0.0}

\newcommand{\righttriangle}{\ding{228}\xspace}

\newcommand{\SoBigDataITAck}{European Union - NextGenerationEU - National Recovery and Resilience Plan (Piano Nazionale di Ripresa e Resilienza, PNRR) - Project: “SoBigData.it - Strengthening the Italian RI for Social Mining and Big Data Analytics” - Prot. IR0000013 - Avviso n. 3264 del 28/12/2021\xspace}

\newcommand{\HPCAck}{All the numerical simulations have been realized on the Linux HPC cluster Caliban of the High-Performance Computing Laboratory of the Department of Information Engineering, Computer Science and Mathematics (DISIM) at the University of L’Aquila.\xspace}

\newcommand{\FairEduAck}{''FAIR-EDU: Promote FAIRness in EDUcation institutions'' a project founded by the University of L'Aquila, 2022\xspace}

\newcommand{\EugainAck}{COST Action CA19122 – European Network Balance in Informatics (EUGAIN)\xspace}

\newcommand{\repeatthanks}{\textsuperscript{\thefootnote}}

\begin{document}
%Data-driven Analysis of gender fairness\\ in the higher education landscape
\title{Data-Driven Analysis of Gender Fairness\\ in the Software Engineering Academic Landscape
\thanks{This work is partially supported by \SoBigDataITAck, by \FairEduAck, and by \EugainAck. \HPCAck
}
}
\titlerunning{Data-driven Analysis of gender fairness in the SE Academic Landscape}
% If the paper title is too long for the running head, you can set
% an abbreviated paper title here
%
\author{Giordano d'Aloisio\orcidID{0000-0001-7388-890X}\thanks{These authors contributed equally to the paper} \and Andrea D'Angelo\orcidID{0000-0002-0577-2494}\repeatthanks \and
Francesca Marzi\orcidID{0009-0009-9129-9231}\repeatthanks \and Diana Di Marco \and Giovanni Stilo\orcidID{0000-0002-2092-0213}\thanks{Corresponding Author} \and Antinisca Di Marco\orcidID{0000-0001-7214-9945}}
\authorrunning{d'Aloisio et al.}
% % First names are abbreviated in the running head.
% % If there are more than two authors, 'et al.' is used.
% %
\institute{University of L'Aquila, L'Aquila, Italy\\
\email{
\{giordano.daloisio,andrea.dangelo6\}@graduate.univaq.it\\
diana.dimarco@student.univaq.it\\
\{francesca.marzi,antinisca.dimarco,giovanni.stilo\}@univaq.it}}

\maketitle              % typeset the header of the contribution

\begin{abstract}
Gender bias in education gained considerable relevance in the literature over the years. However, while the problem of gender bias in education has been widely addressed from a student perspective, it is still not fully analysed from an academic point of view. In this work, we study the problem of gender bias in academic promotions (\ie from Researcher to Associated Professor and from Associated to Full Professor) in the informatics (INF) and software engineering (SE) Italian communities. 
In particular, we first conduct a literature review to assess how the problem of gender bias in academia has been addressed so far. 
Next, we describe a process to collect and preprocess the INF and SE data needed to analyse gender bias in Italian academic promotions. Subsequently, we apply a formal bias metric to these data to assess the amount of bias and look at its variation over time. From the conducted analysis, we observe how the SE community presents a higher bias in promotions to Associate Professors and a smaller bias in promotions to Full Professors compared to the overall INF community.

\end{abstract}

\section{Introduction}\label{sec:intro}

%\todo{Motivazione al problema di bias e fairness in ambito accademico. Fare riferimento anche ad articolo Learning Analytics: An Opportunity for Education di Baker}

Nowadays, the problem of \emph{gender bias} has been widely considered and analysed in the literature under several contexts and domains, like health \cite{ruiz1997two}, justice \cite{angwin2016machine}, or education \cite{moss2012science}. Concerning the latter, the problem of gender bias in education gained considerable relevance over the years, and several papers studied this issue from both a technical and sociological point of view  \cite{baker_algorithmic_2021,mengel_gender_2019}. However, most works focus on gender bias in students' education, not considering other relevant contexts \cite{baker2023learning}. In this work, we want to analyze the issue of gender bias in education from the academic point of view by analyzing if there is a gender bias in academic promotions (\ie from Researcher to Associated Professor and from Associated to Full Professor) in the Italian academic context, in Italian informatics (INF) in general and software engineering (SE) in particular.

In particular, we first perform a literature review to assess how the issue of gender bias in academia has been addressed so far. Next, we perform an empirical analysis of gender bias in academic promotions in the Italian informatics (INF) community. We first extract all the needed data from several open repositories and process them to make them suitable for the analysis. Then, by applying a formal bias metric, we show the trend of bias over the years, starting from 2018 to 2022. Finally, we compare the overall trend with the sole software engineering (SE) Italian community highlighting how the trend for the latter exhibits similar behaviour, albeit considerably more biased towards researchers and less biased towards associate professors, compared to the overall INF community. 

%In particular, the scope of our research is detailed by the following research questions:

% \begin{questions}

% \item \rqone

% We conducted a literature review to answer this question. In particular, we carefully examined all relevant papers related to gender bias in academia, both in Italy and other countries, based on specific inclusion and exclusion criteria. 

%For each paper, we analyzed factors such as the research context (i.e. the country where the study was conducted), the objectives, the type of data used (if any), and the analytical approach employed. Our review revealed a lack of formal analysis of gender bias in the Italian (but also more general) context.

% \item \rqtwo

% To answer this question, we perform an empirical analysis of gender bias in academic promotions in the Italian computer science community. We first extract all the needed data from several open repositories and process them to make them suitable for the analysis. Then, by applying a formal bias metric, we show the trend of bias over the years, starting from 2018 to 2022. Finally, we compare the overall trend with the one of the sole software engineering community highlighting how the trend for the latter exhibits similar behaviour, albeit considerably more biased towards researchers and less biased towards associate professors, compared to the overall Computer Science community. 

% \end{questions}

Hence, the main contributions of this work are the following:
\begin{itemize}
    \item We perform a literature review of the most relevant papers addressing the issue of gender bias in academia by also highlighting the main weaknesses of the current approaches (Section \ref{sec:related});
    \item We describe a process to collect and preprocess data useful to assess the amount of gender bias in academic promotions in Italy (Section \ref{sec:experiment});
    \item We depict the trend of gender bias in academic promotions in Italy over the years by relying on a formal bias metric, and we compare the trend of bias of the overall INF Italian community with the sole SE Italian community (Section \ref{sec:results}).
\end{itemize}

The paper concludes in Section \ref{sec:conclusion} which describes some future works and wraps up the paper.
\vspace{-.3cm}
\section{Gender Bias in Classic Academic Systems}\label{sec:related}

This section describes the literature review process, focused on those works that address the problem of gender bias in academia. The search process involved research of conference proceedings and journal papers on Google Scholar by relying on the search string shown in listing \ref{lst:searchString}.

\begin{lstlisting}[label=lst:searchString,caption=Search string,style=searchstringstyle]
allintitle: gender bias OR academic recruitment OR gender discrimination OR  Women's faculty recruitment OR faculty equity OR career advancements OR Italian universities OR selection processes  
\end{lstlisting}

% \emph{gender bias}, \emph{academic recruitment}, \emph{gender discrimination}, \emph{Women's faculty recruitment}, \emph{faculty equity}, \emph{career advancements}, \emph{Italian universities}, \emph{selection processes}.

Among the results, we selected papers that studied and analysed gender bias in the context of Italian educational systems. Papers discussing practices and techniques utilised in foreign universities were also included to gain a broader perspective and compare different approaches and methods. 
We mainly focus on works related to the recruitment, promotion and productivity level of academic staff, i.e., full professors, associate professors and researchers. Articles about specific faculties or that address the gender bias problem in the general working world are excluded. This process yields 21 papers that have been carefully analysed to highlight these main features: the \textit{context} (i.e. the country where the study was conducted), the \textit{process} (\ie recruitment, promotions or productivity) in which the gender bias has been studied, if the data used are \textit{public} or not, the \textit{analytical method} employed (\ie whether descriptive or inferential statistics are used to analyze the data), and the \textit{year} of the paper.%in which the study has been conducted.

Table \ref{tab:table} summarises such features for each paper. Note that papers with the same features have been grouped in the same row.

\vspace{-.5cm}
\begin{table}
	\centering
	\caption{Summary of the Literature Review.}
	\label{tab:table}
	\resizebox{0.75\textwidth}{!}{
		\begin{tabular}{|c|c|c|c|c|c|}
			\hline
			\textbf{Paper}                                                                & \textbf{Context} & \textbf{Process} & \textbf{Source Data} & \textbf{Analytical Method } & \textbf{Year} \\ \hline
			\textbf{\cite{Todd-2000}}                                                     & AU               & Prom.            & Priv.                & Descr.                      & 2000          \\ \hline
			\textbf{\cite{https://doi.org/10.1111/1467-9485.00199}}                       & U.K.             & Prom.            & Priv.                & Inf.                        & 2001          \\ \hline
			\textbf{\cite{SONNAD2002415}}                                                 & U.S.             & Recr./Prom.      & Priv.                & Inf./Descr.                 & 2002          \\ \hline
			\textbf{\cite{fox2005gender}}                                                 & U.S.             & Prod.            & Priv.                & Inf./Descr.                 & 2005          \\ \hline
			\textbf{\cite{brink}}                                                         & NL               & Recr.            & Priv.                & Descr.                      & 2006          \\ \hline
			\textbf{\cite{article}}                                                       & IT               & Prod.            & Pub.                 & Descr.                      & 2009          \\ \hline
			\textbf{\cite{Glass-2010}}                                                    & \textit{UNK}     & Rescr./Prom.     & Priv.                & Inf./Descr.                 & 2010          \\ \hline
			\textbf{\cite{RePEc:clb:wpaper:201106}}                                       & IT               & Prom.            & Pub.                 & Inf./Descr.                 & 2011          \\ \hline
			\textbf{\cite{abramo2016gender,socsci8050160}}                                & IT               & Recr.            & Pub.                 & Inf./Descr.                 & 2016, 2019    \\ \hline
			\textbf{\cite{10.1093/oep/gpx023,doi:10.1080/03075079.2019.1693990}}          & IT               & Prom.            & Pub.                 & Descr.                      & 2017, 2021    \\ \hline
			\textbf{\cite{RePEc:spr:scient:v:115:y:2018:i:2:d:10.1007_s11192-018-2696-8}} & IT               & Prom.            & Pub.                 & Inf.                        & 2018          \\ \hline
			\textbf{\cite{doi:10.1089/jwh.2019.8027,articlè}}                             & U.S.             & Recr./Prom       & Priv.                & Descr.                      & 2020, 2019    \\ \hline
			\textbf{\cite{sekaquaptewa2019evidence}}                                      & U.S.             & Recr.            & Priv.                & Descr.                      & 2019          \\ \hline
			\textbf{\cite{socsci9090163}}                                                 & IT               & Recr./Priv.      & Pub.                 & Descr.                      & 2020          \\ \hline
			\textbf{\cite{calabrese2021female}}                                           & IT               & Prom.            & Priv.                & Inf./Descr.                 & 2021          \\ \hline
			\textbf{\cite{Carlsson1472311}}                                               & IS,NO,SE         & Recr.            & Priv.                & Inf.                        & 2021          \\ \hline
			\textbf{\cite{https://doi.org/10.1111/jasp.12780}}                            & DE,AT,CH         & Prom.            & Priv.                & Inf./Descr.                 & 2022          \\ \hline
			\textbf{\cite{KENNEY2022114358}}                                              & \textit{UNK}     & Prom.            & Priv.                & Descr.                      & 2022          \\ \hline
		\end{tabular}
	}
\end{table}
\vspace{-.5cm}

Concerning the context, most of the papers focus on specific countries, while the rest of them are generic and unrelated to particular academic system. In the table, we use the official national abbreviation to specify each country, while papers with unspecified countries are labeled with \textit{UNK}.

Concerning the process, most papers address the problem of gender bias either in \textit{recruitment} or \textit{promotions}, while only two papers (\ie \cite{fox2005gender,abramo2016gender}) address the issue of gender bias in \textit{productivity}.
Gender bias in recruitment is mainly addressed by providing recommendations, practices, and strategies to minimize the impact of bias and reach gender equity in the recruitment process. 
Instead, the problem of gender bias in academic promotions is mainly addressed by estimating the probability of promotion by looking at the number of female and male academicians across different career stages or focusing on women in university leadership. Finally, the problem of gender bias in productivity is addressed by investigating the causes that lead to lower productivity by women.

Concerning the source data, public data comes mainly from institutional repositories like the \textit{Ministero dell'Università e della Ricerca (MIUR)} (\ie the Italian Ministry of University and Research) and the National Scientific Qualification website (for Italian works)\cite{miur_website,asn_website}.
Private data were instead collected through different methods, for instance
interviews \cite{article,RePEc:spr:scient:v:115:y:2018:i:2:d:10.1007_s11192-018-2696-8}, questionnaires \cite{abramo2016gender} and compilation of surveys \cite{brink,calabrese2021female,sekaquaptewa2019evidence,sekaquaptewa2019evidence,Todd-2000}.
Other papers collected data directly from internal private university databases.

Concerning the analytical methods, papers using classical descriptive analysis typically measure the percentages of males and females across career stages and institutions, means, standard deviations or comparisons using t-tests between men and women.
In addition to these indicators, cross-tables \cite{Glass-2010}, frequency distributions and segregation indexes \cite{socsci9090163} were used. 
Papers that perform inferential statistical analysis use different regressions methods, such as ordinary least squares regressions, multiple logistic regressions, and multilevel logistic regressions. Works like \cite{socsci8050160} use quantitative analysis with the glass ceiling index and the glass door index to measure and compare the effects of gender practices, and \cite{https://doi.org/10.1111/1467-9485.00199} relies on a static discrete-choice model for rank attainment. 

From this review of the existing literature, it is clear how there is an interest in analysing the issue of gender bias in academia. However, some examined works are old, and the reported conclusions may be outdated. Moreover, we have seen a lack of analyses using formal metrics to measure bias, and none of the reported papers analyses the issue of gender bias in academic promotions inside the informatics community (and thereby software engineering). In this paper, we aim to overcome these lacks by formally analysing gender bias in academic promotions in the informatics (and software engineering) Italian communities.

\section{Analysis Description}\label{sec:experiment}

This section presents the analysis conducted to evaluate the level of gender bias in the academic positions within the overall informatics (INF) and software engineering (SE) Italian communities. The informatics community is the conjunction of Areas 1 and 9 of the MIUR scientific areas classification. \cite{miur}
We first report the dataset creation and filtering procedure. Next, we describe the performed experiment.
\vspace{-.3cm}
\subsection{Data Collection and Filtering}

\begin{figure}[h!]
    \centering
    \includegraphics[width=.65\textwidth]{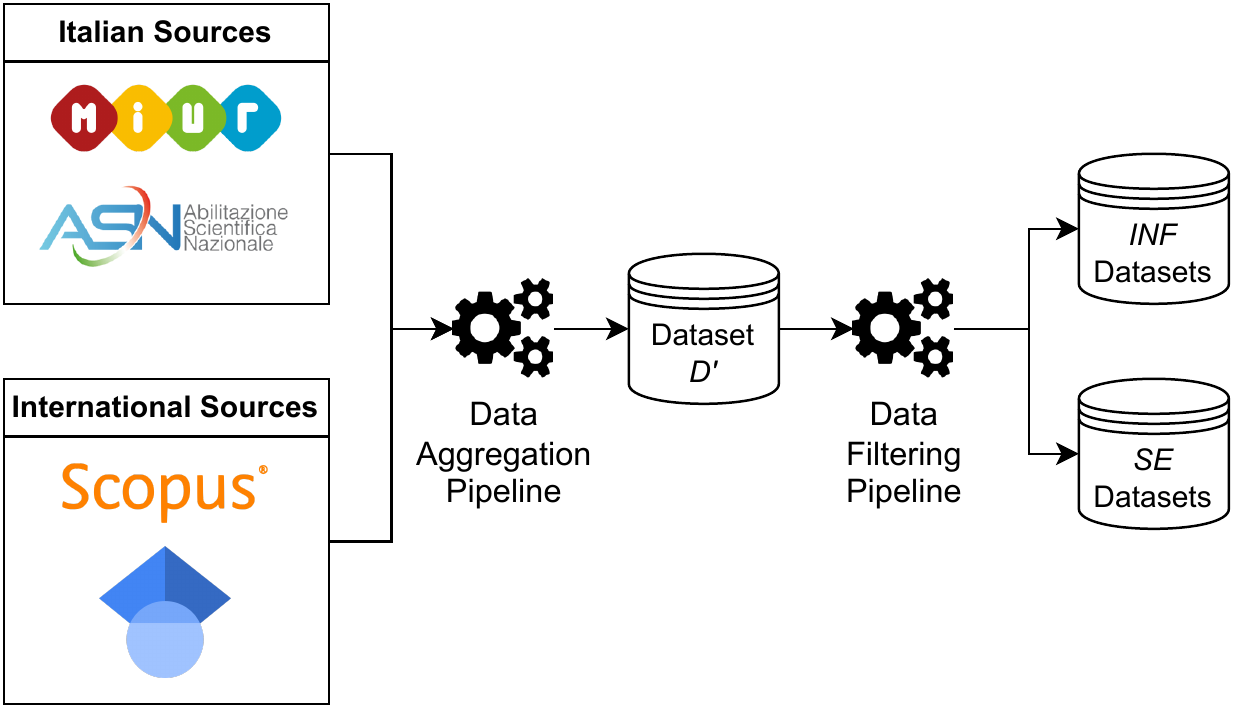}
    \caption{Data collection and filtering pipeline}
    \label{fig:pipeline}
\end{figure}

Figure \ref{fig:pipeline} reports the full data collection and filtering pipeline used to collect the datasets of the INF and SE Italian communities for our analysis\footnote{The source code will be released in the camera-ready version of the paper}. 
In the figure, we report the different sources (Italian and international) where we gathered the needed information.  

The first step of the pipeline is the dataset collection and aggregation. Specifically, data was gathered between 2015 and 2022 with the aim of identifying the following information:  

\noindent \ding{228} \textbf{Personal Data:} \ie information such as age and gender. These data have been gathered from the MIUR website, which contains all the information about people employed in the Italian academia \cite{miur_website}.

\noindent \ding{228} \textbf{Academic Career:} \ie information such as the university and department of affiliation, career advancements, academic seniority, macro disciplinary area, scientific sub-sector they belong to, area of expertise,  current  academic appointment, academics managerial appointments, teaching activities, funded projects, committees, salaries, and sabbatical period. These data have been gathered from the MIUR and National Scientific Qualification (ASN) websites \cite{asn_website}.

\noindent \righttriangle \textbf{Scientific Productivity:} \ie information such as the list of publications, the total number of papers, total citations, the h-index, publication range, papers per year, citations per year, publication types, journal metrics, and research area. These data have been scraped from Scopus \cite{scopus} and Google Scholar \cite{scholar}.

Note that not all the reported information is used in the following analysis, but we choose to gather them  for future works.
The data have then been aggregated into a single dataset $D'$ using the \textit{name}, \textit{surname}, \textit{email}, and \textit{affiliation} as join keys.
This aggregated dataset $D'$ was then thoroughly anonymized to protect the University employees' privacy. As a result, no references to names, surnames, or other sensitive or personal data are stored, as they are neither relevant nor valuable for computing bias metrics. This collected dataset can not be publicly released for legal reasons, however, it can be recreated by gathering the same data from the sources mentioned above.

\begin{figure}[h]
    \centering
    \includegraphics[width=.7\textwidth]{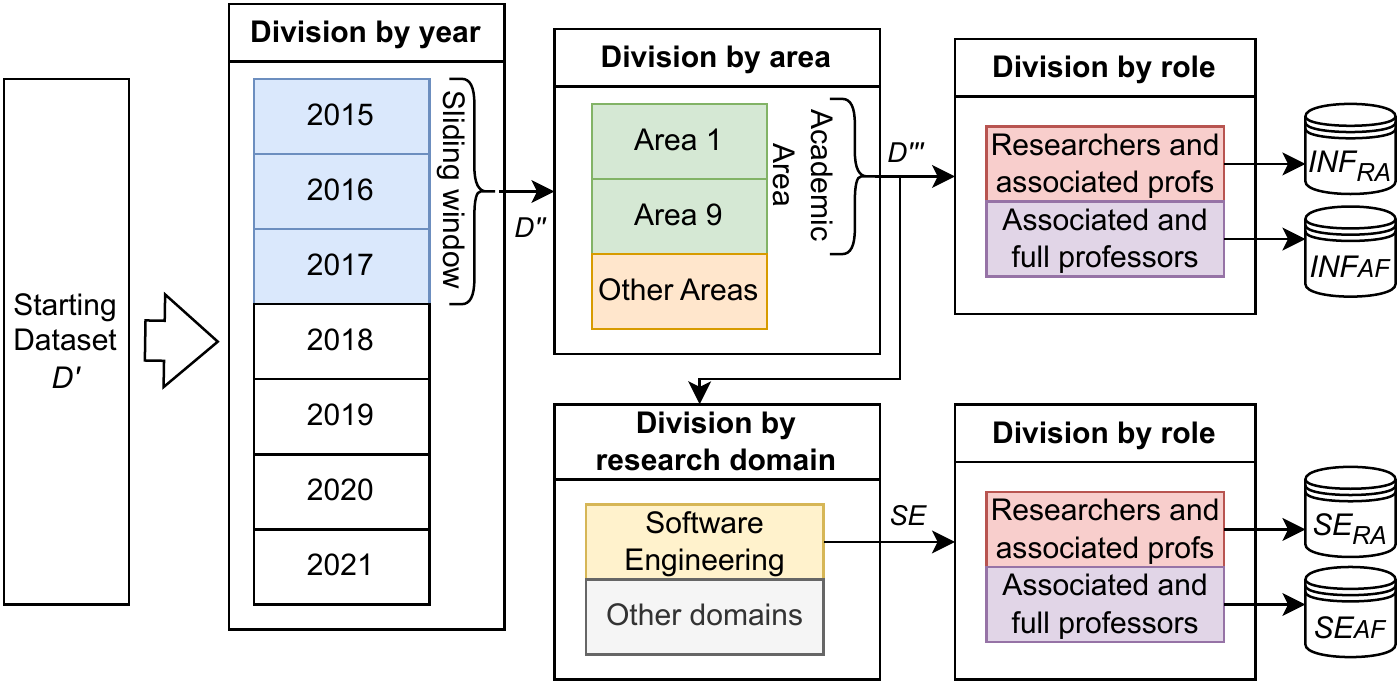}
    \caption{Filtering pipeline of the dataset.}
    \label{img:dataset-pipeline}
    \vspace{-1.0em}
\end{figure}

Starting from the anonymized dataset $D'$, we performed a set of filtering operations to obtain the final datasets that we used to compute bias metrics yearly. The filtering procedure is depicted in Figure \ref{img:dataset-pipeline}.
Since we are interested in the evolution of bias in academic promotions year by year, the anonymized dataset $D'$ was split according to a sliding time window of fixed size. In particular, we considered a sliding window of three years, starting from 2015. 
Hence, to gather metrics for 2019, with the sliding window size set to 3, we would slice $D'$ to obtain only the columns referencing data collected from 2016 to 2019. %This allowed us to obtain, visualize, and analyze a time series of bias metrics from which we could recognize the evolution of the data and draw apt conclusions. 
After this operation, we obtain a partially filtered dataset $D''$ for each sliding window.

The subsequent step was selecting only specific scientific areas from $D''$. Because different domains have different promotion criteria, it would be incorrect to consider them all together. Our study only focused on Areas 1 and 9 of the MIUR scientific areas classification, which refers broadly to Science, technology, engineering, and mathematics \cite{miur}. In this study, we refer to the conjunction of these two areas as the Informatics community. From this further filtering, we obtain a dataset $D'''$. From $D'''$, we perform two different branches of operations.
In the first branch, $D'''$ is split into two versions: one without records representing researchers ($INF_{AF}$) and one without Full Professors ($INF_{RA}$). %The former will be the final dataset for the computation of the bias existing within the context of promotion from researcher to Associate Professor ($CS_{RA}$). At the same time, the latter will be used to compare Associate Professors with Full Professors ($CS_{AF}$). 
In the second phase, $D'''$ is refined by selecting individuals who work specifically in the SE field. To achieve this, we use Google Scholar to find individuals who have expressed interest in \textit{software engineering} or related topics such as \textit{software architecture}, \textit{model-driven engineering}, \textit{software quality}, and \textit{software testing}. The SE dataset is then divided into two sub-datasets as done above: one consisting of only researchers and associate professors ($SE_{RA}$), and the other consisting of only associate and full professors ($SE_{AF}$).
As a result of the data pre-processing pipeline, four distinct datasets were created. Two of them are for the overall Italian INF community ($INF_{RA}$ and $INF_{AF}$), while the other two are for the Italian SE community (${SE}_{RA}$ and ${SE}_{AF}$). Finally, we only preserved data for workers  employed at an Italian university for the entire time window.

\subsection{Analysis Setting}

Once the final yearly datasets $INF_{RA}$, $INF_{AF}$, ${SE}_{RA}$, and ${SE}_{AF}$ have been constructed, the experiments can occur.
As already mentioned, the experiment aims to measure the amount of gender bias in academic promotions and analyze its variation over the years. To calculate the amount of bias, we use the \textit{Disparate Impact (DI)} metric \cite{feldman_certifying_2015}. This metric measures the probability of having a \textit{positive outcome} while being in the \textit{privileged} or \textit{unprivileged} group and is defined formally as: 

\begin{equation}    
DI = \frac{P(Y=y_p|X=x_{unpriv})}{P(Y=y_p|X=x_{priv})}
\label{def:DI}
\end{equation}
where $Y$ is the label, $y_p$ is the positive outcome, $X$ is the sensitive variable, and $x_{unpriv}$ and $x_{priv}$ are the values identifying the unprivileged and privileged groups, respectively. The more this metric is close to one, the fairer the dataset.

In our context, the label assigned to a person represents their position for that particular year.
%, which could be a \textit{Researcher}, an \textit{Associate Professor}, or a \textit{Full Professor}. 
In the analysis between Researchers and Associate Professors, the positive label is \textit{Associate Professor}, while in the analysis between Associate and Full Professors, it is \textit{Full Professor}. The sensitive variable is \textit{gender}, where \textit{men} and \textit{women} are the privileged and unprivileged groups, respectively.  Hence, the experiment is performed as follows: for each final yearly dataset ($INF_{RA}$, $INF_{AF}$, ${SE}_{RA}$, and ${SE}_{AF}$) and for each year in the considered range (2018-2022), we compute the DI between the two subsets contained in the dataset (either Researchers and Associate Professors or Associate Professors and Full Professors). We also compute the cardinality of each subset per year.

\section{Experimental Results}\label{sec:results}

\begin{figure}[h]
    \centering
    \includegraphics[width=.9\textwidth]{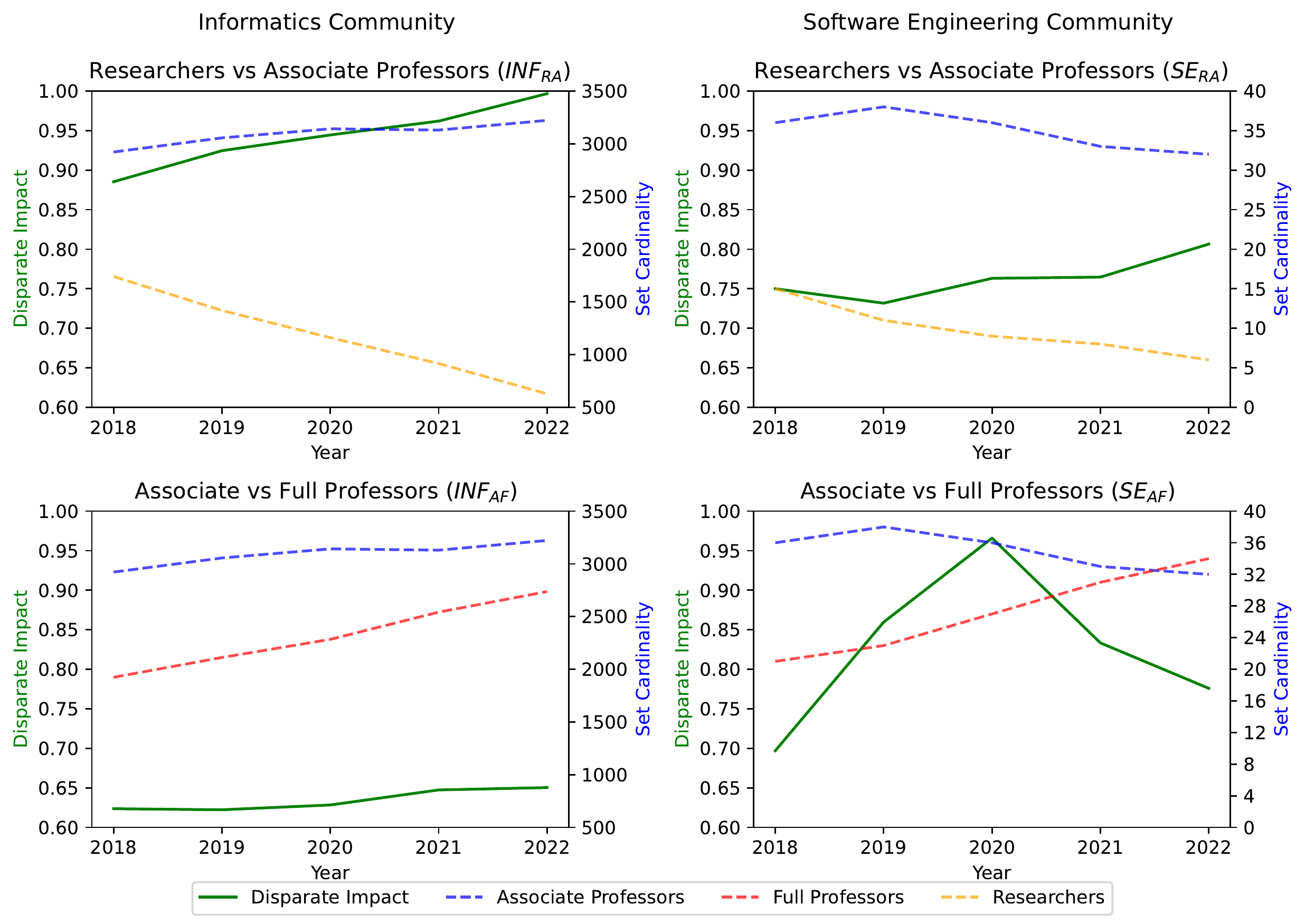}
    \caption{Year-by-year Disparate Impact and Set Cardinality for the Informatics Community (left column) and Software Engineering Community (right column).}
    \label{img:grid_result}
\end{figure}

In this section, we present and discuss the Experimental Results. Figure \ref{img:grid_result} shows the Disparate Impact (DI) (left y-axis) and set cardinalities (right y-axis) for each of the datasets above ($INF_{RA}$, $INF_{AF}$, ${SE}_{RA}$, and ${SE}_{AF}$) on a yearly basis in the reference period (2018-2022). In the figure, the charts on the left side show results for the Informatics (INF) Community datasets ($INF_{RA}$, $INF_{AF}$), while the ones on the right side show results for the Software Engineering (SE) Community ($SE_{RA}$, $SE_{AF}$).

Concerning the full set cardinalities (\ie of both men and women), they exhibit the same trend across all datasets. Since we only consider people that were in the Italian academic system for the entire reference period, we do not consider researchers that were acquired later than 2018, so their cardinality is bound to decrease. The number of Full professors is rising in both the INF and SE communities, but the increase in the SE community is significantly larger. In 2022, there are more Full professors than Associate professors specifically in the SE subset. This suggests that promotions to Full professorship are occurring at a higher rate among professors in the field of SE compared to the INF community.

Concerning the gender bias in promotions to Associate Professor ($INF_{RA}$ and $SE_{RA}$ in the figure), in both the Informatics and Software Engineering communities the trend of Disparate Impact (DI) appears to be on an upward trajectory. However, the SE community seems to suffer from a higher bias w.r.t. the overall INF community. The DI for the SE community starts from a value of 0.75 in 2018 to a value of 0.8 in 2022. In contrast, the DI of the INF community starts from a value of 0.9 in 2018 to a value of almost 1 in 2022, meaning a nearly complete absence of bias in academic promotions. In general, we observe how the amount of bias in the SE community is about 20\% higher than in the overall INF community.

In contrast, concerning bias in promotions to Full Professors ($INF_{AF}$ and $SE_{AF}$ in the figure), the SE community exhibits a much lower bias concerning the INF community. DI for the SE community starts from 0.7 in 2018, then reaches a peak of 0.95 in 2020, to a final value of almost 0.8 in 2022. This downtrend from 2020 to 2022 can be partially explained by the small set cardinality, which makes the DI more sensitive to small changes (\ie additions or deletions) in the groups. Instead, the DI for the overall INF community presents a slight increase over the period, starting from a value of 0.63 in 2018 to a value of 0.65 in 2022. In this case, the amount of bias in the INF community ranges from 15 to 35\% greater than in the SE community throughout the observed period.

\vspace{-.2cm}
\section{Conclusion and Future Work}\label{sec:conclusion}
% \vspace{-.2cm}
In this paper, we have studied the issue of gender bias in academic promotions. First, we performed a literature review to observe how the literature has addressed this issue  so far. Then, we formally analyzed gender bias in academic promotions in the informatics (INF) and software engineering (SE) Italian communities. From the analysis, we observed that gender bias has been improving over the years in both communities, even though the SE community has a higher trend in promoting professors from Associated to Full compared to the broader INF community. In the future, we plan to extend this analysis to other countries by identifying valuable data sources to retrieve all the needed information. Next, we plan to analyze the behaviour of a Machine Learning classifier trained on such data to predict the position of a person. In particular, we want to study how a classifier is subject to learning a possible gender bias in the data and how we can mitigate it by relying on proper fairness methods.

\bibliographystyle{splncs04}
\bibliography{bibliography}

\end{document}